\documentclass[doublecol]{epl2}   %for 2 columns style without line numbers
\usepackage{bm}
\usepackage{amsfonts}
\usepackage{latexsym}
\usepackage[latin1]{inputenc}
\usepackage{graphicx}
\usepackage{amsmath}
\usepackage{palatino}
\usepackage{mathpazo}
\usepackage{textcomp}
\usepackage{float}
\usepackage{booktabs}
\usepackage{dcolumn}
\usepackage{hyperref}
\hypersetup{colorlinks,citecolor=blue}
\usepackage{amsmath}
\usepackage{xcolor}

\title{Embedding procedure and wormhole solutions in $f(Q)$ gravity}
\shorttitle{Embedding procedure and wormhole solutions in $f(Q)$ gravity} %Insert here a short version of the title if it exceeds 70 characters

\author{Zinnat Hassan\inst{1}\footnote{E-mail: zinnathassan980@gmail.com}, G. Mustafa\inst{2}\footnote{E-mail: gmustafa3828@gmail.com}, Joao R.L. Santos\inst{3}\footnote{E-Mail: joaorafael@df.ufcg.edu.br} \and P. K. Sahoo\inst{1}\footnote{E-mail: pksahoo@hyderabad.bits-pilani.ac.in}}
\shortauthor{Z. Hassan \etal}

\institute{                    
  \inst{1} Department of Mathematics, Birla Institute of Technology and
Science-Pilani,\\ Hyderabad Campus, Hyderabad-500078, India.\\
 \inst{2} Department of Physics, Zhejiang Normal University,
Jinhua, 321004, People's Republic of China.\\
\inst{3} UFCG - Universidade Federal de Campina Grande - Unidade Acad\^{e}mica de F\'isica,  58429-900 Campina Grande, PB, Brazil.}
\abstract{
An intriguing solution that appears in General Relativity (GR) but has not been observed so far is the wormhole. This exotic solution describes a topological bridge connecting two distinct universes or two different points in the same universe. It is known that the traversable wormhole solutions violate all the energy conditions in GR, resulting in their instability. In this work, we are going to unveil new wormhole solutions for $f(Q)$ gravity where $Q$ is the non-metricity scalar, which is responsible for the gravitational interaction. The energy conditions to constraint these wormhole solutions were derived using the embedding procedure. This procedure consists of rewriting the density and the pressures of the solutions as those presented by General Relativity. Then, the nontrivial contributions coming from new theories of gravity are embedded into the effective equations for density and pressures. Along with our approach, we carefully analyze two families of $f(Q)$ models and we used two different shape functions to build the wormholes solutions for each of these $f(Q)$ models. We are going to present new scenarios with the possibility of traversable wormholes satisfying SEC or DEC energy conditions in the presence of exotic matter.}

\begin{document}

\maketitle

\section{Introduction}

Today we are facing one of the most exciting times in astrophysics and cosmology. Experiments such as, Laser Interferometer Gravitational-Wave Observatory (LIGO) \cite{Gourgoulhon/2019}, Event Horizon Telescope (EHT) \cite{Chael/2016, Akiyama/2019}, Virgo \cite{Abuter/2020}, 
%International Gamma-Ray Astrophysics Laboratory (INTEGRAL) \cite{Winkler/2003}, 
Advanced Telescope for High-Energy Astrophysics (ATHENA) \cite{Barcons/2017}, 
%Imaging x-ray Polarimetry mission (IXPE) \cite{Soffitta/2013} 
and Swift \cite{Burrows/2005}, have been responsible for changing the bars of our knowledge about the nature of gravity and astrophysical objects. They have also been used to set new constraints over different theories of gravity, selecting viable models among the enormous amount of proposals presented in the literature.\\
An exotic type of solution that appears in General Relativity but has not been observed so far is the wormhole (WH) \cite{Visser/1995}. The wormhole is a topological bridge connecting points in two distinct universes or two different points in the same universe. Although they have never been observed, several investigations in the literature propose the viability of finding wormholes. One possible strategy of observation was recently proposed by Bueno et al. \cite{Bueno/2018}, using gravitational waves measurements to study echoes of the gravitational wave signal at the horizon scale of black holes. These gravitational waves would be related to the postmerger ringdown phase in binary coalescences. Another interesting idea for observation was introduced by Paul et al. \cite{Paul/2020}, using distinctive features of the accretion disk images to distinguish between a wormhole geometry and a black hole.\\
Beyond the observations, it is also relevant to a search for traversable wormholes, which means wormholes that are big enough that a person could traverse them and survive the tidal forces. This class of wormholes was proposed in the seminal paper of Morris and Thorne \cite{Morris/1988} and is still under investigation, as we can see in the recent manuscript of Maldacena and Milekhin \cite{Maldacena/2021}.\\ 
%There the authors propose a new dark sector based on the Randall-Sundrum model \cite{RS/1999} with a $U(1)$ gauge field, where traversable wormholes could exist. This dark sector interacts with the Standard Model particles only through gravity.
The main issue related to traversable wormholes is that these solutions violate all energy conditions in General Relativity. Therefore we are impelled to search for traversable wormhole solutions in different theories of gravity that could obey energy conditions, allowing their existence.\\
As it is known, wormholes can also be used as prototypes to test several theories of gravity \cite{Capozziello/2011,Capozziello/2015}. 
%One advantage of searching wormholes in modified gravity is that we can find an alternative route to understand the exotic matter, which will enable the existence of traversable solutions. 
%In the past decades several investigations were presented looking for wormholes in $f(R)$ \cite{Lobo/2009, Lobo/2009/01, Lobo/2009/02, Lobo/2009/03, Lobo/2009/04}, and $f(R,T)$ \cite{Moraes/2019, Moraes/2019/01} gravity theories.\\
A new and promising theory of gravity is the so-called $f(Q)$ gravity, introduced by Jimenez et al. \cite{Jimenez/2018}, where the nonmetricity $Q$ is responsible for the gravitational interaction. In the last few years, several observational data have been tested in $f(Q)$ gravity. Some interesting constraints over $f(Q)$ gravity were proposed by Lazkoz et al. \cite{Lazkoz/2019}. There the authors tested the viability of $f(Q)$ gravity using data from the expansion rate, Type Ia Supernovae, Quasars, Gamma-Ray Bursts, Baryon Acoustic Oscillations data, and Cosmic Microwave Background distance. Moreover, the viability of $f(Q)$ gravity models in respect to energy conditions was successfully proven by Mandal et al. \cite{Mandal/2020}, where the authors introduced an embedding procedure to include non-trivial contributions of the nonmetricity function into the energy constraints.\\
In this work, we apply the embedding procedure to derive traversable wormhole solutions in $f(Q)$ gravity. This procedure consists of rewriting the density and the pressures of the solutions as those presented by General Relativity. Then, the nontrivial contributions coming from new theories of gravity are embedded into the effective equations for density and pressures. The advantage of such a methodology is that we can find the energy conditions using the standard constraints from the Raychaudhuri equation. The embedding procedure allowed us to study the null, dominant, and strong energy conditions for wormholes in the presence of an anisotropic fluid.  We carefully analyze two families of $f(Q)$ models, and we also used two different shape functions,  a $r^{\,n}$ and $r\, e^{1-\frac{r}{r_0}}$ ones, to build the wormholes solutions for each of these $f(Q)$ models. It is relevant to point that some of us had recently reported new traversable wormhole families of solutions in $f(Q)$ \cite{Mustafa/2021}. However, the energy constraints previously determined in \cite{Mustafa/2021} did not consider the embedding procedure, so the current work presents an updated approach for finding traversable wormholes compatible with investigations recently introduced on cosmological constraints for $f(Q)$, and $f(Q,T)$ gravity \cite{Mandal/2020, Arora/2021}. Along with our discussions, we are going to present scenarios with the possibility of traversable wormholes, satisfying at least one set of energy conditions in the presence of exotic matter.\\
%The ideas presented in this manuscript are organized as follows: In sec-2, we approached the generalities about the $f(Q)$ gravity. In sec-3 we investigated wormhole solutions in $f(Q)$ gravity, the embedding procedure, and we presented the energy conditions. Then, in sec-4 we analyzed the energy conditions for two $f(Q)$ families, considering two different sets of shape functions for the traversable wormholes. Our final remarks and perspectives for further works are presented in sec-5.

\section{The $f(Q)$ gravity}\label{sec2}

In this section, we are going to briefly present some generalities about the $f(Q)$ gravity. The so-called symmetric teleparallel gravity or $f(Q)$ gravity was introduced by Jimenez et al. \cite{Jimenez/2018}. The action for this modified gravity is given by 
\begin{equation}\label{1}
\mathcal{S}=\int\frac{1}{2}\,f(Q)\sqrt{-g}\,d^4x+\int \mathcal{L}_m\,\sqrt{-g}\,d^4x\, ,
\end{equation}
where $g$ is the determinant of the metric tensor $g_{\mu\nu}$, and $\mathcal{L}_m$ is Lagrangian density of matter.\\
The nonmetricity tensor is given by\\
\begin{equation}\label{2}
Q_{\lambda\mu\nu}=\bigtriangledown_{\lambda} g_{\mu\nu}=\partial_\lambda g_{\mu\nu}-\Gamma^\beta_{\,\,\,\lambda \mu}g_{\beta \nu}-\Gamma^\beta_{\,\,\,\lambda \nu}g_{\mu \beta},
\end{equation}
where $\Gamma^\beta_{\,\,\,\mu\nu}$ is the affine connection and it can be decomposed into the following components,
\begin{equation}
\Gamma^\beta_{\,\,\,\mu\nu}=\lbrace^\beta_{\,\,\,\mu\nu} \rbrace+K^\beta_{\,\,\,\mu\nu}+ L^\beta_{\,\,\,\mu\nu},
\end{equation}
where $\lbrace^\beta_{\,\,\,\mu\nu} \rbrace$ is the Levi-Civita connection defined by
\begin{equation}
\lbrace^\beta_{\,\,\,\mu\nu} \rbrace=\frac{1}{2}g^{\beta\sigma}\left(\partial_\mu g_{\sigma\nu}+\partial_\nu g_{\sigma\mu}-\partial_\sigma g_{\mu\nu}\right),
\end{equation}
the disformation $L^\beta_{\,\,\,\mu\nu}$ is defined by
\begin{equation}
L^\beta_{\,\,\,\mu\nu}=\frac{1}{2}Q^\beta_{\,\,\,\mu\nu}-Q_{(\mu\,\,\,\,\,\,\nu)}^{\,\,\,\,\,\,\beta},
\end{equation}
and $K^\beta_{\,\,\,\mu\nu}$ is the contortion written as
\begin{equation}
K^\beta_{\,\,\,\mu\nu}=\frac{1}{2} T^\beta_{\,\,\,\mu\nu}+T_{(\mu\,\,\,\,\,\,\nu)}^{\,\,\,\,\,\,\beta},
\end{equation}
with the torsion tensor $T^\beta_{\,\,\,\mu\nu}$ defined as the anti-symmetric part of the affine connection, $T^\beta_{\,\,\,\mu\nu}=2\Gamma^\lambda_{\,\,\,[\mu\nu]}$.\\
One can also establish a superpotential related to the nonmetricity tensor as
\begin{eqnarray}\label{4} \nonumber
P^\alpha\;_{\mu\nu}&=&\frac{1}{4}\left[-Q^\alpha\;_{\mu\nu}+2Q_{(\mu}\;^\alpha\;_{\nu)}+Q^\alpha g_{\mu\nu}\right.\\
&&
\left.-\tilde{Q}^\alpha g_{\mu\nu}-\delta^\alpha_{(\mu}Q_{\nu)}\right],
\end{eqnarray}
where
\begin{equation}
\label{3}
Q_{\alpha}=Q_{\alpha}\;^{\mu}\;_{\mu},\; \tilde{Q}_\alpha=Q^\mu\;_{\alpha\mu}.
\end{equation}
are two independence traces.\\
which enable us to define the nonmetricity scalar as
\begin{equation}
\label{5}
Q=-Q_{\alpha\mu\nu}\,P^{\alpha\mu\nu}.
\end{equation}

In order to find the field equations for this theory of gravity, we can set that the action \eqref{1} is constant in respect to variations over the metric tensor $g_{\mu\nu}$, resulting in
\begin{multline}\label{7}
\frac{2}{\sqrt{-g}}\bigtriangledown_\gamma\left(\sqrt{-g}\,f_Q\,P^\gamma\;_{\mu\nu}\right)+\frac{1}{2}g_{\mu\nu}f \\
+f_Q\left(P_{\mu\gamma i}\,Q_\nu\;^{\gamma i}-2\,Q_{\gamma i \mu}\,P^{\gamma i}\;_\nu\right)=-T_{\mu\nu},
\end{multline}
where $f_Q=\frac{df}{dQ}$, and $T_{\mu\,\nu}$ is the standard energy-momentum tensor, whose form is
\begin{equation}\label{6}
T_{\mu\nu}=-\frac{2}{\sqrt{-g}}\frac{\delta\left(\sqrt{-g}\,\mathcal{L}_m\right)}{\delta g^{\mu\nu}}.
\end{equation}

Moreover, by taking action \eqref{1} constant in respect to variations over the connection, we are able to derive the extra constraint
\begin{equation}\label{8}
\bigtriangledown_\mu \bigtriangledown_\nu \left(\sqrt{-g}\,f_Q\,P^\gamma\;_{\mu\nu}\right)=0.
\end{equation}

The torsionless and curvatureless make the affine connection $\Gamma$ have the following form \cite{Jimenez/2018}
\begin{equation}\label{i}
\Gamma^\lambda_{\,\,\,\mu\nu}=\left(\frac{\partial x^\lambda}{\partial\xi^\beta}\right)\partial_\mu \partial_\nu \xi^\beta.
\end{equation}
where $\xi$ is the frame of transformation. This transformation of $\Gamma^\lambda_{\,\,\,\mu\nu}$ is the core of the covariant formulation of $f(Q)$ gravity, which preserves the Lorentz covariance under the nonmetricity framework. Two excellent reviews about this calculation were nicely presented by Xu et al. in \cite{Xu/2019}, and by Jimenez et al. \cite {Koivisto}\\
In this study, we have considered a special coordinate choice, the so-called coincident gauge, so that $\Gamma^\lambda_{\,\,\,\mu\nu}=0$.\\
Then, the nonmetricity \eqref{2} reduces to
\begin{equation}
Q_{\lambda\mu\nu}=\partial_\lambda g_{\mu\nu},
\end{equation}
and thereby vastly simplifies the calculation since only the metric is the fundamental variable. However, in this case, the action no longer remains diffeomorphism invariant, except for standard General Relativity \cite{Koivisto}. One can use the covariant formulation of $f(Q)$ gravity to avoid such an issue. Since the affine connection in Eq. \eqref{i} is purely inertial, one could use the covariant formulation by first determining affine connection in the absence of gravity \cite{Zhao}.
\section{Wormholes in $f(Q)$ gravity}
\label{sec3}

%Once we would like to find wormholes solutions for $f(Q)$ gravity, 
Let us consider the standard static spherically symmetric Morris and Thorne metric in Schwarzschild coordinates \cite{Morris/1988}, given by
\begin{eqnarray}\label{9} \nonumber
ds^2&=&-e^{2\Phi(r)}dt^2+\left(1-\frac{b(r)}{r}\right)^{-1}dr^2 \\ 
&&
+r^2d\theta^2+r^2\text{sin}^2\theta d\phi^2\,,
\end{eqnarray}
where $b(r)$, and $\Phi(r)$ denote the shape function and redshift function, respectively. Moreover, the radial coordinate $r$ varies between $r_0\leq r<\infty$,  with $r_0$ as the radius of the throat. To avoid the presence of event horizons, the redshift function $\Phi(r)$ is must be finite everywhere. Besides, the shape function $b(r)$ must obey the following conditions to ensure a traversable wormhole:

\begin{itemize}
\item Throat condition: at the throat i.e. at $r=r_0$, the shape function must satisfy the constraint $b(r_0)=r_0$, and for $r>r_0$ we have $1-\frac{b(r)}{r}>0$.
\item Flaring out condition: $b^{\,\prime}(r_0)<1$, at wormhole's throat.
\item Asymptotically flatness condition: $\frac{b(r)}{r}\rightarrow o$ as $r\rightarrow \infty$.
\end{itemize}

In this study, we will work with the energy-momentum tensor for an anisotropic fluid which is define by
\begin{equation}\label{10}
T_{\mu}^{\nu}=\left(\rho+P_t\right)u_{\mu}\,u^{\nu}+P_t\,\delta_{\mu}^{\nu}+\left(P_r-P_t\right)v_{\mu}\,v^{\nu},
\end{equation}
where $\rho$ and $u_{\mu}$ are the density and the four-velocity vector, respectively. This energy-momentum tensor was introduced by Letelier \cite{Letelier/1980} to describe a two fluid model in plasma physics, and has been applied in several distinct scenarios so far, as for instance, in models for magnetized neutron stars \cite{Debabrata/2021}. Besides $v_{\mu}$ is the unitary space-like vector in the radial direction, $P_r$ is the radial pressure in the direction of $u_{\mu}$, and $P_t$ is the tangential pressure orthogonal to $v_{\mu}$.\\
Using Eq. \eqref{5}, we can determine the nonmetricity scalar for the metric \eqref{9} is given by
\begin{equation}\label{14}
Q=-\frac{2}{r}\left(1-\frac{b(r)}{r}\right)\left(2\phi{'}(r)+\frac{1}{r}\right).
\end{equation}
So, by taking the previous energy-momentum tensor together with the metric \eqref{9}, into the field equations \eqref{7}, we obtain
\begin{equation}
\label{11}
\frac{b^{'}}{r^2}=-\frac{1}{f_Q}\left(\rho-\frac{f}{2}\right)+\frac{r-b}{r^2}\left[\frac{f_{QQ}}{f_Q}\,Q^{'}+2\phi^{'}+\frac{1}{r}\right],
\end{equation}
\begin{equation}
\label{12}
-\frac{b}{r^3}=-\frac{1}{2\,f_Q}\left(p_r+\frac{f}{2}\right)-\frac{1}{2\,r^2}\left[1+4(r-b)\phi^{'}\right],
\end{equation}
\begin{multline}
\label{13}
-\frac{rb^{'}-b}{2r^3}=-\frac{1}{f_Q}\left(P_t+\frac{f}{2}\right)-\\
\frac{r-b}{r^2}\left[\frac{f_{QQ}}{f_Q}\,Q^{'}+\frac{1}{1+r\,\phi{'}}\left(\frac{1}{r}+\phi{'}(3+r\,\phi{'})+r\,\phi{''}\right)\right],
\end{multline}
where ${'}$ represents $\frac{d}{dr}$.\\
Since the redshift function $\phi(r)$ should be finite and non-vanishing at the WH throat. So we have considered $\phi(r)=constant$ in this current study and hence $\phi{'}(r)=0$.

\subsection{The Embedding Procedure}
In order to successfully embed the non-trivial contributions from $f(Q)$ gravity into the energy density and pressures related to the traversable wormhole solution, let us rewrite Eqs. \eqref{11} - \eqref{13} in the following forms
%\begin{widetext}
\begin{multline}\label{15}
\rho=\frac{1}{r}\left(1-\frac{b}{r}\right) \left[2f_{QQ}Q^{'}+f_Q\left(\frac{1}{r}-\frac{b^{'}}{r-b}\right)+\right.\\ \left.f\,\frac{r^2}{2(r-b)}\right],
\end{multline}
\begin{equation}
\label{16}
p_r=\frac{1}{r}\left(1-\frac{b}{r}\right) \left[f_Q\left(\frac{2\,b-r}{r(r-b)}\right)-f\,\frac{r^2}{2(r-b)}\right],
\end{equation}
\begin{multline}\label{17}
p_t=\frac{1}{2\,r}\left(1-\frac{b}{r}\right) \left[-2\,f_{QQ}Q^{'}+f_Q\,\left(\frac{r\,b^{'}+b-2\,r}{r(r-b)}\right)-\right.\\ \left.f\,\frac{r^2}{2(r-b)}\right].
\end{multline}
%\end{widetext}
%Besides, as its known, the Morris-Throne traversable wormhole equations for General Relativity \cite{Francisco/2005} are
%\begin{equation}
%\label{18}
%\frac{b^{'}(r)}{r^2}=\tilde{\rho}\,,
%\end{equation}
%\begin{equation}
%\label{19}
%-\frac{b(r)}{r^3}=\tilde{P_r}\,,
%\end{equation}
%\begin{equation}
%\label{20}
%-\frac{rb^{'}(r)-b(r)}{2r^3}=\tilde{P_t}\,.
%\end{equation}
Comparing Eqs. \eqref{11} - \eqref{13} with the Morris-Throne traversable wormhole equations for General Relativity \cite{Francisco/2005}, we are able to find the following effective density and pressures
\begin{equation}
\label{21}
\tilde{\rho}=\frac{1}{r}\left(1-\frac{b}{r}\right)\left(\frac{1}{r}+\frac{2\,f_{QQ}Q^{'}}{f_Q}\right)-\frac{1}{f_Q}\left(\rho-\frac{f}{2}\right)\,,
\end{equation}
\begin{equation}
\label{22}
\tilde{P_r}=-\frac{1}{2}\left(\frac{1}{r^2}+\frac{1}{f_Q}\left(P_r+\frac{f}{2}\right)\right)\,,
\end{equation}
\begin{equation}
\label{23}
\tilde{P_t}=-\frac{1}{r}\left(1-\frac{b}{r}\right)\left(\frac{1}{r}+\frac{f_{QQ}Q^{'}}{f_Q}\right)-\frac{1}{f_Q}\left(P_t+\frac{f}{2}\right)\,.
\end{equation}
These previous relations unveil that the nonmetricity contributions are embedded into $\tilde{\rho}$, $\tilde{P_r}$, and  $\tilde{P_t}$, and are crucial to properly determine the different set of energy conditions for the wormholes. 
%This embedding procedure introduced by Mandal et al. \cite{Mandal/2020}, was used to constraint $f(Q)$ gravity through energy conditions in the cosmological context. The procedure was also successfully applied to determine viable cosmological $f(Q,T)$ models \cite{Arora/2021}. 

\subsection{Energy Conditions}

The energy conditions are fundamental tools for modified theories of gravity, since they enable us to carefully analyze the casual and geodesic structure of space-time. One path to derive such conditions is through the Raychaudhuri equations, which describe the action of congruence and attractiveness of the gravity for timelike, spacelike, or lightlike curves.
%As it is known, the Raychaudhuri equations for General Relativity are written as \cite{Raychaudhuri/1955, Raychaudhuri/1955/01, Raychaudhuri/1955/02}
%\begin{equation}\label{1a}
%\frac{d\theta}{d\tau}-\omega_{\mu\nu}\,\omega^{\mu\nu}+\sigma_{\mu\nu}\sigma^{\mu\nu}+\frac{1}{3}\theta^2+R_{\mu\nu}u^\mu\,u^\nu=0\,,
%\end{equation}
%\begin{equation}\label{2a}
%\frac{d\theta}{d\tau}-\omega_{\mu\nu}\,\omega^{\mu\nu}+\sigma_{\mu\nu}\sigma^{\mu\nu}+\frac{1}{2}\theta^2+R_{\mu\nu}\eta^\mu\eta^\nu=0 .
%\end{equation}
%Here, $\eta^\mu$, $\theta$, $\sigma^{\mu\nu}$ and $\omega_{\mu\nu}$ are the null vector, the expansion factor, the shear and the rotation associated with the vector field $u^\mu$, respectively. As we expect an attractive gravity regime, Eqs. \eqref{1a} and \eqref{2a} yields to the constraints
%\begin{equation}\label{3a}
%R_{\mu\nu}u^\mu\,u^\nu\geq0\,,
%\end{equation}
%\begin{equation}\label{4a}
%R_{\mu\nu}\eta^\mu\eta^\nu\geq0\,,
%\end{equation}
%respectively. 
Since we are working with an anisotropic fluid, the energy conditions for standard GR are such that

$\bullet$ Strong energy conditions (SEC) if $\tilde{\rho}+\tilde{P_j}\geq0$, $\tilde{\rho}+\sum_j\tilde{P_j}\geq0$, $\forall j$.\\
$\bullet$ Dominant energy conditions (DEC) if $\tilde{\rho}\geq0$, $\tilde{\rho}\pm \tilde{P_j}\geq0$, $\forall j$.\\
$\bullet$ Weak energy conditions (WEC) if $\tilde{\rho}\geq0$, $\tilde{\rho}+\tilde{P_j}\geq0$, $\forall j$.\\
$\bullet$ Null energy condition (NEC) if $\tilde{\rho}+\tilde{P_j}\geq0$, $\forall j$.\\
where $j=r,t$ and $P_j$ and $\rho$ denotes the pressures and energy density, respectively.\\

So, taking the effective density and pressures from Eqs. \eqref{21}-\eqref{23}, we yield to
%\begin{widetext}
\begin{multline}
\label{28}
\tilde{\rho}+\tilde{P_r}=-\frac{1}{2\,f_Q}\left(\rho+P_r\right)+\frac{1}{2r}\left(-\frac{2}{r}+\frac{r\,b^{'}+b}{r^2}\right)\\
+\frac{1}{r}\left(1-\frac{b}{r}\right)\left(\frac{1}{r}+\frac{f_{QQ}Q^{'}}{f_Q}\right);
\end{multline}
\begin{equation}
\label{29}
\tilde{\rho}+\tilde{P_t}=-\frac{1}{f_Q}\left(\rho+P_t\right)+\frac{1}{r}\left(1-\frac{b}{r}\right)\frac{f_{QQ}Q^{'}}{f_Q};
\end{equation}
\begin{multline}
\label{30}
\tilde{\rho}-\tilde{P_r}=-\frac{1}{2\,f_Q}\left(\rho-P_r\right)+\frac{f}{2\,f_Q}+\frac{r\,b^{'}+b}{2\,r^3}\\
+\frac{1}{r}\left(1-\frac{b}{r}\right)\left(\frac{1}{r}+\frac{f_{QQ}Q^{'}}{f_Q}\right);
\end{multline}
\begin{eqnarray}
\label{31}
&& 
\tilde{\rho}-\tilde{P_t}= -\frac{1}{2\,f_Q} \\ \nonumber
&&
\times\left(\rho-P_t\right)+\frac{f}{f_Q}+\frac{1}{r}\left(1-\frac{b}{r}\right)\left(\frac{2}{r}+\frac{3\,f_{QQ}Q^{'}}{f_Q}\right);
\end{eqnarray}
\begin{eqnarray}
\label{32}\nonumber
&&
\tilde{\rho}+\tilde{P_r}+2\tilde{P_t}= -\frac{1}{2\,f_Q}\\ 
&&
\times \,\left(\rho+P_r+2\,P_t\right)+\frac{b}{r^3}-\frac{f}{2\,f_Q}.
\end{eqnarray}
Therefore, the previous equations enable us to compute the following set of energy conditions for wormholes at $f(Q)$ gravity:

WEC: $\,\,\,\,\, \tilde{\rho}\geq 0\,\, \Rightarrow \,\,\,\,\,\, \rho\geq0 \,\,\,$ with $\,\,f_Q\leq0$, $\,\,\frac{1}{r}\left(1-\frac{b}{r}\right)\left(\frac{1}{r}+\frac{2\,f_{QQ}Q^{'}}{f_Q}\right)+\frac{f}{2\,f_Q}\geq0.$ \\
NEC: $\,\,\,\,\, \tilde{\rho}+\tilde{P_r}\geq 0\,\, \Rightarrow \,\,\,\,\,\, \rho+P_r\geq0 \,\,\,$ with $f_Q\leq0$,$\,\,\,\,\,\frac{1}{2r}\left(-\frac{2}{r}+\frac{r\,b^{'}+b}{r^2}\right)+\frac{1}{r}\left(1-\frac{b}{r}\right)\left(\frac{1}{r}+\frac{f_{QQ}Q^{'}}{f_Q}\right)\geq0\,\,\,\,$ and $\,\,\, \tilde{\rho}+\tilde{P_t}\geq 0\,\, \Rightarrow \,\,\,\,\ \rho+P_t\geq0 \,\,\,$ with $\,\,\,f_Q\leq0$,$\,\,\,\frac{1}{r}\left(1-\frac{b}{r}\right)\frac{f_{QQ}Q^{'}}{f_Q}\geq0$.\\
DEC: $\,\,\,\,\, \tilde{\rho}-\tilde{P_r}\geq 0\,\, \Rightarrow \,\,\,\,\,\, \rho-P_r\geq0 \,\,\,$ with $\,\,\,f_Q\leq0$,$\,\,\,\,\frac{r\,b^{'}+b}{2\,r^3}+\frac{1}{r}\left(1-\frac{b}{r}\right)\left(\frac{1}{r}+\frac{f_{QQ}Q^{'}}{f_Q}\right)\geq0\,\,\,\,$ and $\,\,\,\,\, \tilde{\rho}-\tilde{P_t}\geq 0\,\, \Rightarrow \,\,\,\,\,\, \rho-P_t\geq0 \,\,\,$ with $\,\,\,f_Q\leq0$,$\,\,\,\,\frac{f}{f_Q}+\frac{1}{r}\left(1-\frac{b}{r}\right)\left(\frac{2}{r}+\frac{3\,f_{QQ}Q^{'}}{f_Q}\right)\geq0$.\\
SEC: $\,\,\,\,\, \tilde{\rho}+\tilde{P_r}+2\tilde{P_t}\geq 0\,\, \Rightarrow \,\,\,\,\,\, \rho+P_r+2P_t\geq0 \,\,\,$ with $\,\,\,f_Q\leq0$,$\,\,\,\,\frac{b}{r^3}-\frac{f}{2\,f_Q}\geq0$.

\section{Wormholes in $f(Q)$ Gravity}
\label{sec4}

After presenting the generalities and the constraints over the energy conditions, we can test the viability of different families of $f(Q)$ gravity to properly describe stable wormhole solutions. In this section, we present our analyzes considering two specific forms of $f(Q)$ models, (i) the non-linear model, and (ii) the logarithmic  model.
\subsection{Model-I: $f(Q)=\alpha Q+\beta Q^n$}
Let us firstly consider a non-linear form of $f(Q)$, i.e., $f(Q)=\alpha Q+\beta Q^n$ \cite{Solanki/2022}, with $\alpha$, $\beta$, and $n$ are free parameters. In this article, we study with the particular form of this model by considering $n=4$.
%These free parameters are going to be constrained through the energy conditions, and $\beta$ parameter plays an analogous role of vacuum energy or a cosmological constant. 
%This specific $f(Q)$ model was introduced by Harko et al. \cite{Harko/2018}, where the authors studied the viability of cosmological scenarios for $f(Q)$, including coupling between the nonmentricity scalar and matter. 
%Moreover, an analogous model in $f(Q,T)$ gravity was constrained by Arora et al. \cite{Arora/2020}, using Supernovae data sets.
In order to build wormhole solutions, we are going to work with the shape functions $b(r)=r_0\left(\frac{r}{r_0}\right)^m$, and $b(r)=r\,e^{1-\frac{r}{r_0}}$. The first shape function was presented in the seminal work of Lobo et al. \cite{Lobo/2009}, where it was verified the viability of wormhole solutions for different $f(R)$ gravity theories. The second shape function was introduced by Moraes et al. \cite{Moraes_2/2019}, in order to fulfill the energy conditions constraints for wormholes in $f(R,T)$ gravity theories. Moreover, these shape functions satisfies all the basic criteria of traversable wormhole.

\subsection{Shape function (SF-1): $b(r)=r_0\left(\frac{r}{r_0}\right)^m$}
\begin{figure}[htbp]
\centerline{\includegraphics[width=3.5in, height=2in]{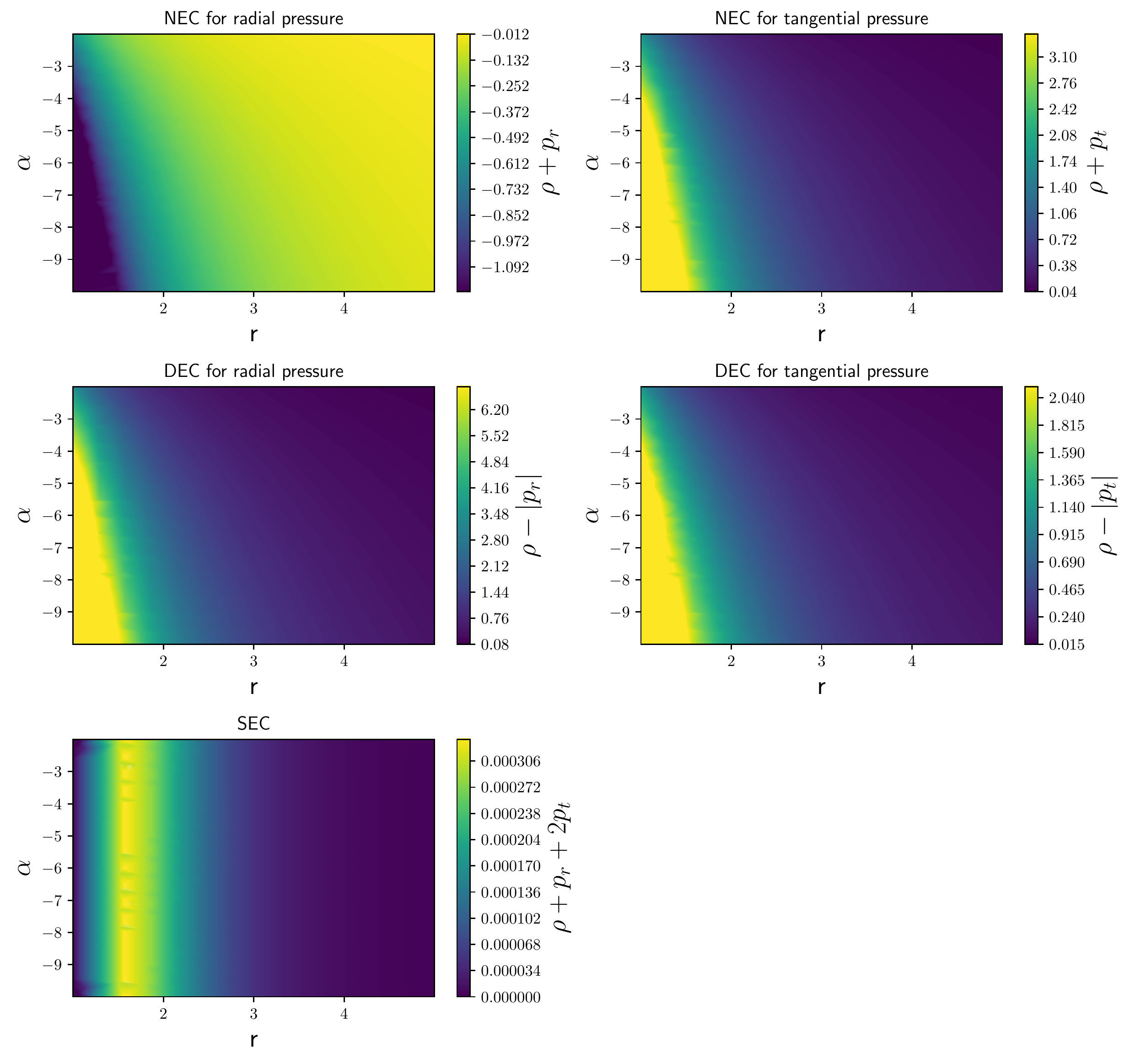}}
\caption{Evolution of energy conditions with $n=4,\,\,\beta=0.8\,\,m=0.7$ and $r_0=1$ for the shape function $b(r)=r_0\left(\frac{r}{r_0}\right)^n$.}
\label{fig1}
\end{figure}

By taking this $f(Q)$ gravity family, as well as the shape function $b(r)=r_0\left(\frac{r}{r_0}\right)^n$ into Eqs. \eqref{15}-\eqref{17}, we get the following density and pressures,
%\begin{widetext}
\begin{eqnarray}
\label{33}
&& \nonumber
\hspace{-0.8 cm}\rho =\frac{\beta  2^{n-1} \left(\frac{r_0 \left(\frac{r}{r_0}\right)^m-r}{r^3}\right)^n}{r-r_0 \left(\frac{r}{r_0}\right)^m} \left(r_0 (n (2 n (m-3)-m+7)-1)\right.\\ 
&&
\hspace{-0.8 cm}\left. \left(\frac{r}{r_0}\right)^m+(n-1) (4 n-1) r\right)-\frac{\alpha  m r_0 \left(\frac{r}{r_0}\right)^m}{r^3},
\end{eqnarray}
\begin{eqnarray}
\label{34}
&& \nonumber
\hspace{-0.8 cm}
P_r=\frac{\beta  2^{n-1} \left(\frac{r_0 \left(\frac{r}{r_0}\right)^m-r}{r^3}\right)^n}{r-r_0 \left(\frac{r}{r_0}\right)^m} \left((n-1) r-(2 n-1)r_0 \left(\frac{r}{r_0}\right)^m\right)\\
&&
\hspace{-0.8 cm}+\frac{\alpha r_0 \left(\frac{r}{r_0}\right)^m}{r^3},
\end{eqnarray}
\begin{multline}
\label{35}
P_t=\frac{\alpha  (m-1)r_0 \left(\frac{r}{r_0}\right)^m}{2\,r^3}-\\
\frac{\beta\,2^{n-2}(2 n-1) \left(\frac{r_0 \left(\frac{r}{r_0}\right)^m-r}{r^3}\right)^n }{r-r_0 \left(\frac{r}{r_0}\right)^m} \left(r_0 (n (m-3)+2) \left(\frac{r}{r_0}\right)^m \right.\\ \left. +2 (n-1) r\right).
\end{multline}
Now, combining Eqs. \eqref{33}-\eqref{35} with the constraints found at the end of section \ref{sec3}, we have shown the behavior of energy conditions which is depicted in Fig. \ref{fig1}. In the last section, we will discuss the behavior of our obtained graphs.
%\end{widetext}
%\begin{figure}[H]
%\centering
%\includegraphics[scale=0.34]{sp1a.pdf}
%\caption{Characteristic of shape function $b(r)=r_0\left(\frac{r}{r_0}\right)^n$ with $n=0.4$ and $r_0=0.75$ (for model-I). In this case $b(r)-r$ cuts the $r$ axis at $r=0.75$.}
%\label{a}
%\end{figure}
%In Fig. \ref{a} we carefully analyze the behavior of $b(r)$, $b^{'}(r)$, $\frac{b(r)}{r}$ and $b(r)-r$. The values for free parameters $n$ and $r_0$ were chosen to derive an asymptotically flatness behavior, and to satisfy the flaring out condition related to the shape function. One can see from Fig. \ref{a} that the shape function $b(r)$ increases as $r$ gets bigger. Moreover, the asymptotically flatness condition is also satisfied as $\frac{b(r)}{r}\rightarrow 0$ in the limit $r\rightarrow \infty$. Also we find that $b^{'}(r)<1$ at $r=r_0$, meaning that the flaring out condition is obeyed. 
\subsection{Shape function (SF-2): $b(r)=r\,e^{1-\frac{r}{r_0}}$}
\begin{figure}[htbp]
\centerline{\includegraphics[width=3.5in, height=2in]{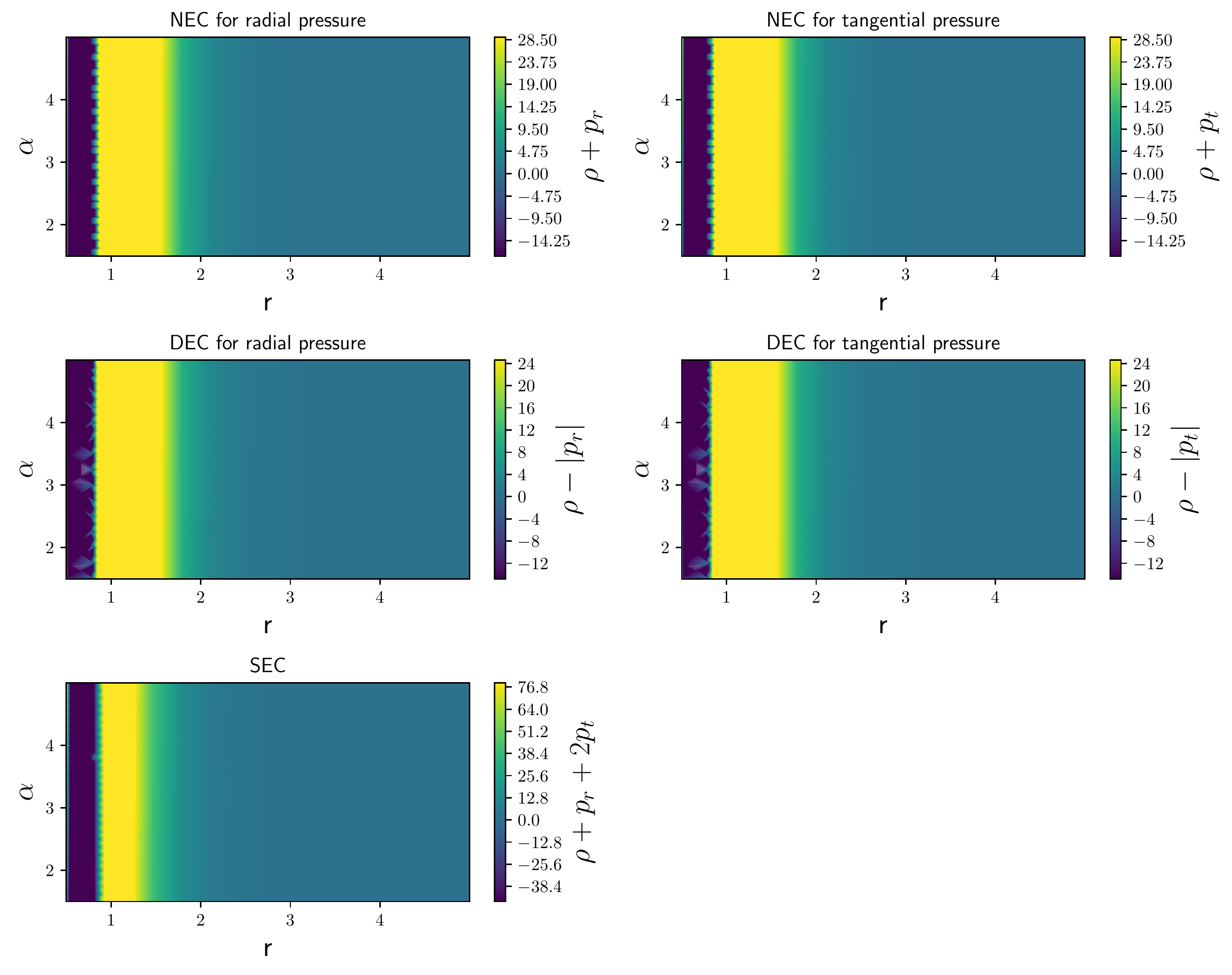}}
\caption{Evolution of energy conditions with $\beta=5,\,\,n=4$ and $r_0=0.5$ for the shape function $b(r)=r\,e^{1-\frac{r}{r_0}}$.}
\label{fig2}
\end{figure}

As a next example, let us work with the quadratic form of $f(Q)$ together with the exponential shape function $b(r)=r\,e^{1-\frac{r}{r_0}}$.  So, by taking these ingredients into Eqs. \eqref{15}-\eqref{17} we derive our density and pressures for these wormhole solutions as follows
%\begin{widetext}
\begin{multline}
\label{36}
\rho=\frac{\beta  2^{n-1} \left(\frac{e^{1-\frac{r}{r_0}}-1}{r^2}\right)^n}{r_0(e^{r/r_0}-e)}\left((n-1) (4 n-1)\,r_0 e^{r/r_0}-\right.\\ \left. e \left(2 n^2 (r+2\,r_0)-n (r+6 r_0)+r_0\right)\right)+\frac{\alpha  e^{1-\frac{r}{r_0}} (r-r_0)}{r_0\,r^2},
\end{multline}
\begin{eqnarray}
\label{37}
&& \nonumber
P_r=\frac{\beta  2^{n-1} \left((n-1) e^{r/r_0}-2 e n+e\right) \left(\frac{e^{1-\frac{r}{r_0}}-1}{r^2}\right)^n}{e^{r/r_0}-e} \\ 
&&
+\frac{\alpha  e^{1-\frac{r}{r_0}}}{r^2},
\end{eqnarray}
\begin{eqnarray}
\label{38}
&& \nonumber
P_t=\frac{\beta  2^{n-1} \left((n-1) e^{r/r_0}-2 e n+e\right) \left(\frac{e^{1-\frac{r}{r_0}}-1}{r^2}\right)^n}{e^{r/r_0}-e}\\
&&
+\frac{\alpha  e^{1-\frac{r}{r_0}}}{r^2}.
\end{eqnarray}
The behavior of energy conditions have been shown in Fig. \ref{fig2} by using the above equations along with the constraints found in section-3.
%By repeating the procedure of our previous analyses, we can substitute Eqs. \eqref{36}-\eqref{38} into the constraints found in section-3, to determine the energy conditions for these wormhole solutions. In Fig. \ref{fig2}, we present the  behavior of the energy conditions derived from $\rho$, $P_t$, and $P_r$.
%\end{widetext} 
%In order to constraint the free parameter $r_0$, we depicted Fig. \ref{c}. There, taking $r_0=0.75$, we find that the shape function satisfies all the basic properties required for a wormhole solution. 
%\begin{figure}[H]
%\centering
%\includegraphics[scale=0.34]{sp2a.pdf}
%\caption{Features of shape function $b(r)=r\,e^{1-\frac{r}{r_0}}$ with $r_0=0.75$ (for model-I).}
%\label{c}
%\end{figure}

%\begin{figure}[H]
%\includegraphics[scale=0.35]{Model-2.pdf}
%\caption{Evolution of energy conditions with $\alpha=0.01$ and $r_0=1.5$ for the shape function $b(r)=r\,e^{1-\frac{r}{r_0}}$.}
%\label{fig2}
%\end{figure}
%\begin{figure}[H]
%\centering
%\includegraphics[scale=0.34]{fig12.pdf}
%\caption{Evolution of energy density $\rho+P_r+2P_t$ with $\alpha=0.01$ and $r_0=0.75$ for the shape function $b(r)=r\,e^{1-\frac{r}{r_0}}$.}
%\label{fig8}
%\end{figure}

%By repeating the procedure of our previous analyses, we can substitute Eqs. \eqref{36}-\eqref{38} into the constraints found in section-3, to determine the energy conditions for these wormhole solutions. In Figs \ref{fig5}-\ref{fig7}, we present the  behavior of the energy conditions derived from $\rho$, $P_t$, and $P_r$.
%
%The figure shows that the energy density is positive, i.e., $\rho>0$, throughout the spacetime. Besides, one can be seen that NEC is violated at wormhole throat due to its radial pressure for any positive $\beta$ and very small range of $r$, configuring the presence of exotic matter at wormhole throat, as discussed in our first example. Moreover, we observe that DEC is satisfied  while the SEC is violated, corroborating with $b(r)=r_0\left(\frac{r}{r_0}\right)^n$ case.  

\subsection{Model-II: $f(Q)=Q+\alpha\,Q^2+\beta\,Q^2 Log(\beta\,Q)$}

As a second path to find wormhole solutions in $f(Q)$ gravity, we carefully investigated a logarithmic form of $f(Q)$, i.e., $f(Q)=Q+\alpha\,Q^2+\beta\,Q^2 Log(\beta\,Q)$. This model is analog to the $R^2$ Starobinsky inflation model \cite{a,b}, which may clarify both the early-time and the late-time acceleration cycles. In $f(Q)$ gravity, this model combines several nonlinear nonmetricity functions, generalizing the $f(Q) = Q^{m}$ family introduced by Mandal et al. \cite{Mandal/2020}. Here $\alpha$ and $\beta$ are the free parameters which are going to be constrained through the energy conditions.\\
Analogously to Model-I, the wormhole solutions are determined using the shape functions $b(r)=r_0\left(\frac{r}{r_0}\right)^n$ and  $b(r)=r\,e^{1-\frac{r}{r_0}}$.

\subsection{Shape function (SF-1): $b(r)=r_0\left(\frac{r}{r_0}\right)^m$}
\begin{figure}[htbp]
\centerline{\includegraphics[width=3.5in, height=2in]{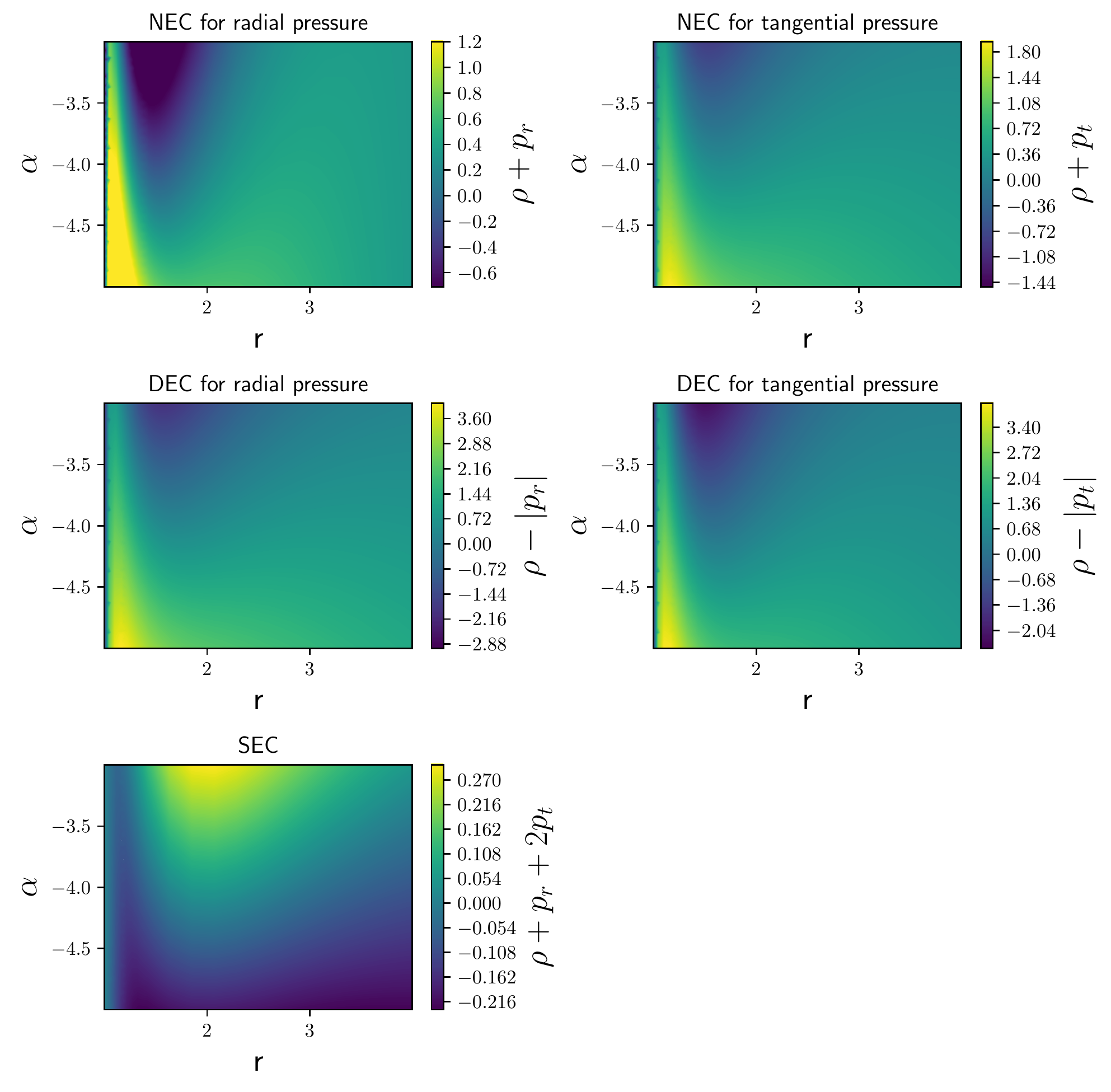}}
\caption{Behavior of energy conditions with $\beta=3,\,\,\,m=2$ and $r_0=1$ for the shape function $b(r)=r_0\left(\frac{r}{r_0}\right)^m$.}
\label{fig3}
\end{figure}

By taking the present $f(Q)$ gravity family, together with the shape function $b(r)=r_0\left(\frac{r}{r_0}\right)^n$ into Eqs. \eqref{15}-\eqref{17}, we were able to derive analytic expressions for $\rho$, $P_r$, and $P_t$ which we omitted for the sake of simplicity. The graphical behavior of the energy conditions derived through these parameters are depicted in Fig. \ref{fig3}.

\subsection{Shape function (SF-2): $b(r)=r\,e^{1-\frac{r}{r_0}}$}
\begin{figure}[h]
\centerline{\includegraphics[width=3.5in, height=2in]{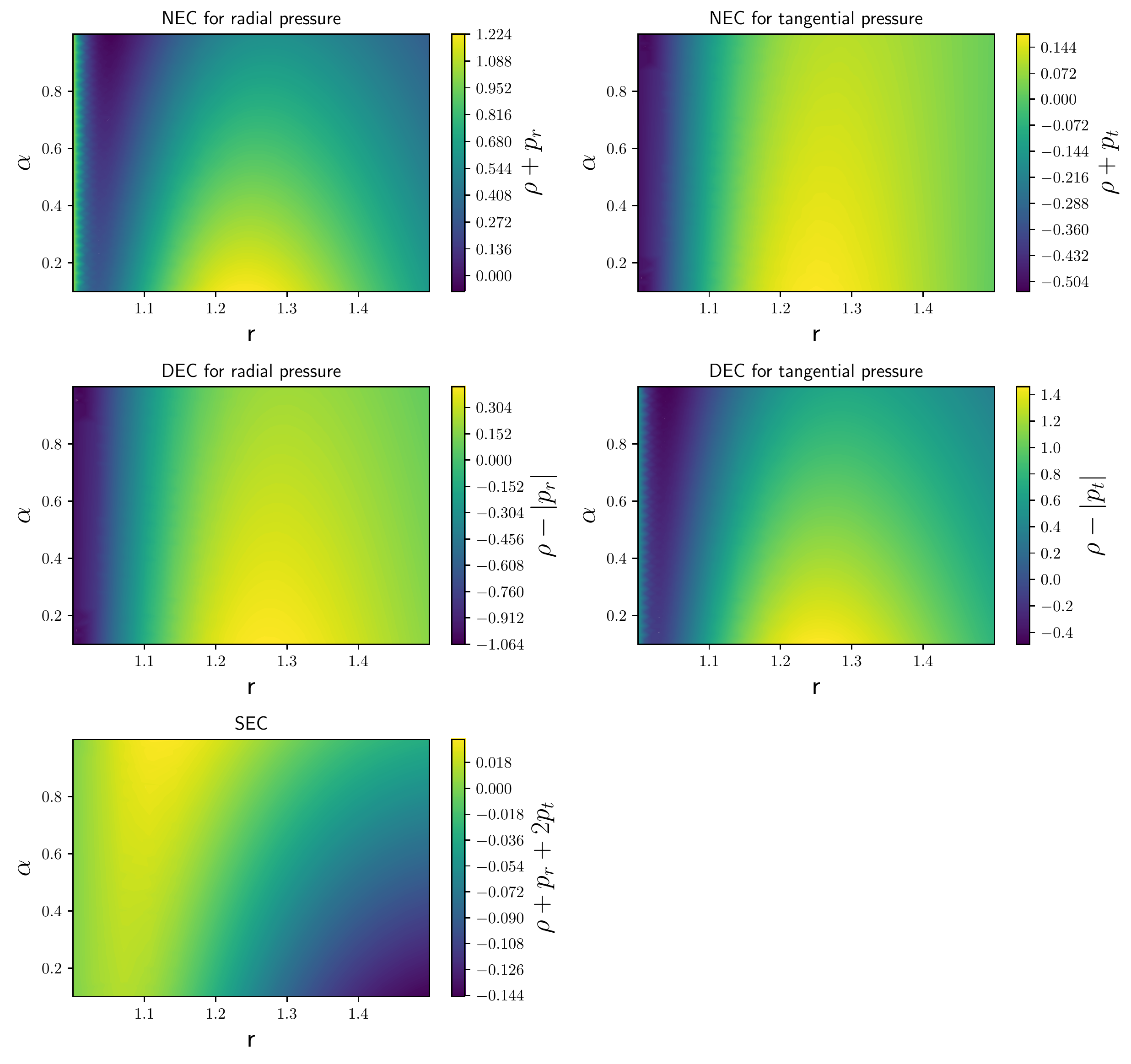}}
\caption{Behavior of energy conditions with $\beta=-2$ and $r_0=1$ for the shape function $b(r)=r\,e^{1-\frac{r}{r_0}}$.}
\label{fig4}
\end{figure}

As a final example, let us work with the exponential form of $f(Q)$ together with the exponential shape function $b(r)=r\,e^{1-\frac{r}{r_0}}$.  Thus, by taking them into Eqs. \eqref{15}-\eqref{17} we were able to derive analytic forms for density and pressures $P_r$ and $P_t$ which were suppressed for simplicity. The energy conditions obtained using these parameters are depicted in Fig. \ref{fig4}.
%\end{widetext}
%\begin{figure}[H]
%\centering
%\includegraphics[scale=0.34]{sp2b.pdf}
%\caption{Features of shape function $b(r)=r\,e^{1-\frac{r}{r_0}}$ with $r_0=1.3$ (for model-II).}
%\label{d}
%\end{figure}

\section{Results and discussions}
\label{sec5}

This letter carefully investigated the viability of wormholes solutions in $f(Q)$ gravity, considering specific sets of shape functions, and an energy-momentum tensor for an anisotropic fluid.\\
In order to find the energy conditions, we made use of the constraints imposed by the Raychaudhuri Eq. using the embedding procedure introduced by Mandal et al. \cite{Mandal/2020}. The constraints derived together with the restrictions over the shape functions allowed us to fix our free parameters, yielding to wormholes solutions that satisfy at least one energy condition for positive energy density. The present methodology using the embedding procedure also complements the discussions recent carried by some of us in \cite{Mustafa/2021}.\\
Firstly we derive the energy conditions for model I, considering both shape functions (SF-1 and SF-2). For SF-1, we saw that NEC is violated for the radial pressure for any value of $\beta$ with $m<1$. For $m>1$, NEC will be satisfied. Also, DEC is satisfied in the entire spacetime and SEC is violated for bigger values of $r$. For SF-2, we have found that all the energy conditions are violated close to the throat for $r<1$. Such features are the same presented by wormholes solutions in GR non-minimally coupled with scalar fields. These behaviors indicate the possibility of stable traversable wormhole solutions for $f(Q)$ gravity, with the exotic matter at their throats. \\
For model-II, we consider logarithmic function of $Q$ and studied the energy conditions for two different shape functions. For SF-1, we observed that NEC is violated for both pressures close to the throat, for $\alpha \geq -3.5$ and $m\geq 2$. Besides, we find that DEC is violated in the neighborhood of the throat and SEC is violated for $\alpha \leq -3.5$. Again for SF-2, we found that NEC is violated for the tangential pressure close to the throat if we work with a small range of positive values of $\alpha$ and for negative values of $\beta$. DEC is also violated in the vicinity of the WH throat whereas SEC is satisfied for $\alpha \approx 1$ or for $r$ close to the throat. Wormholes that do not obey DEC were found as solutions of the so-called $f(R,T)$ gravity \cite{Elizalde/2018}. Although, the wormholes derived in \cite{Elizalde/2018} also satisfy NEC, indicating that they do not have exotic matter at their throats. Consequently, our work presents an interesting scenario with the possibility of stable exotic traversable wormhole solutions which satisfy SEC in the presence of exotic matter. The standard General Relativity approach can be recovered if we take $\alpha = \beta = 0$ in both $f(Q)$ families \cite{Jimenez/2018}. In such cases, all energy conditions for traversable wormholes would be violated as pointed by Morris and Thorne \cite{Morris/1988} .

The methodology here presented can be easily applied to new families of $f(Q)$ as well as for several other theories of gravity, opening a new path in deriving wormhole solutions, and to constraint them through energy conditions. Moreover, we could also work with a different equation of state using the Chaplygin gas \cite{Elizalde/2018} or the Van der Waals fluid. It would also be interesting to constraint our free parameters by combining the energy conditions together with another noncommutative geometry, as recently approached by Hassan et al.  \cite{Hassan/2021}.

\acknowledgments Z.H. thanks the Department of Science and Technology (DST), Government of India, New Delhi, for awarding a Senior Research Fellowship (File No. DST/INSPIRE Fellowship/2019/IF190911). JRLS thanks CNPq (Grant nos. 420479/2018-0, and 309494/2021-4), and PRONEX/CNPq/FAPESQ-PB (Grant nos. 165/2018, and 0015/2019). PKS thanks National Board for Higher Mathematics (NBHM) under Department of Atomic Energy (DAE), Govt. of India for financial support to carry out the Research project No.: 02011/3/2022 NBHM(R.P.)/R\&D II/2152 Dt.14.02.2022. The authors also would like to thank the anonymous referee for comments which improved the quality of this work.

\end{document}